\documentclass[aps,prl,twocolumn,showpacs,superscriptaddress]{revtex4-1}
\usepackage{graphicx}
\usepackage{amsmath}
\usepackage{color}

\begin{document}

\title{Engineering Surface Critical Behavior of (2+1)-Dimensional O(3) Quantum Critical Points}

\author{Chengxiang Ding}
\email{dingcx@ahut.edu.cn}
\affiliation{School of Science and Engineering of Mathematics and Physics, Anhui University of Technology, Maanshan, Anhui 243002, China }

\author{Long Zhang}
\email{longzhang@ucas.ac.cn}
\affiliation{Kavli Institute for Theoretical Sciences and CAS Center for Excellence in Topological Quantum Computation, University of Chinese Academy of Sciences, Beijing 100190, China}

\author{Wenan Guo}
\affiliation{Department of Physics, Beijing Normal University, Beijing 100875, China}

\date{\today}

\begin{abstract}
Surface critical behavior (SCB) refers to the singularities of physical quantities on the surface at the bulk phase transition. It is closely related to and even richer than the bulk critical behavior. In this work, we show that three types of SCB universality are realized in the dimerized Heisenberg models at the (2+1)-dimensional O(3) quantum critical points by engineering the surface configurations. The ordinary transition happens if the surface is gapped in the bulk disordered phase, while the gapless surface state generally leads to the multicritical special transition, even though the latter is precluded in classical phase transitions because the surface is in the lower critical dimension. An extraordinary transition is induced by the ferrimagnetic order on the surface of the staggered Heisenberg model, in which the surface critical exponents violate the results of the scaling theory and thus seriously challenge our current understanding of extraordinary transitions.
\end{abstract}
\maketitle 

\emph{Introduction.---}Universality is a central concept in physics, and plays a key role in the study of phase transitions. The universality suggests that critical exponents in spontaneously symmetry-breaking transitions are determined by the broken symmetry and the spatial dimensions. Moreover, the critical exponents in different universality classes obey the same scaling relations.

When a system with boundaries undergoes a phase transition, physical quantities measured on the surface also show singularities with universal behavior. This is called surface critical behavior (SCB) \cite{Binder1983phase}. Approaching the bulk critical point, both the surface and the bulk correlation lengths diverge, and the long-range order also sets in on the surface. Besides its direct relevance to experiments on realistic materials with boundaries, the SCB is also theoretically appealing. Similar to the bulk critical points, the SCB is also classified according to the universal properties, which are characterized by the surface critical exponents. The surface universality classes are closely related to the bulk ones, and are even richer than the latter because of the extra tunability on the surface. In other words, there is a one-to-many correspondence between the bulk and the surface universality classes.

\begin{figure}[!b]
\includegraphics[width=\linewidth]{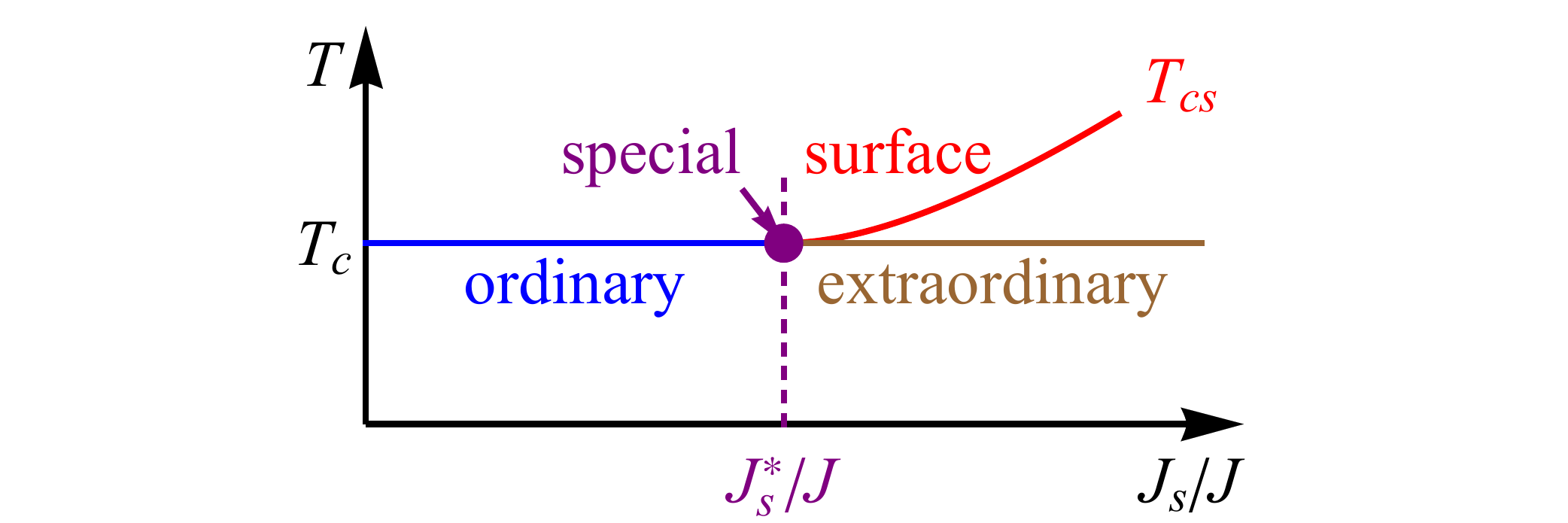}
\caption{Schematic phase diagram of the 3D classical Ising model with boundaries. $J$ and $J_{s}$ are the bulk and the surface coupling strengths, respectively.}
\label{fig:phasediag}
\end{figure}

In classical phase transitions, different surface universality classes can be realized by tuning the surface coupling strength. The phase diagram of the prototypical three dimensional (3D) Ising model is sketched in Fig. \ref{fig:phasediag} \cite{Binder1974, Binder1983phase}. If the coupling in the surface layer $J_{s}$ is comparable to the bulk coupling $J$, the surface remains disordered throughout the bulk disordered phase, thus the surface singularities at the bulk $T_{c}$ are purely induced by the bulk critical state. This is called ``ordinary transition''. If $J_{s}/J\gg 1$, the surface undergoes a 2D phase transition at a higher temperature $T_{cs}>T_{c}$. At the bulk phase transition, the surface exhibits extra singularities, which is called ``extraordinary transition''. The surface $T_{cs}$ and the bulk $T_{c}$ merge at a fine-tuned surface coupling strength $J_{s}^{*}$, where both the surface and the bulk states are critical. This multicritical point is called ``special transition''.

For 3D O($n$) ($n\geq 3$) models, however, the 2D surface alone cannot have O($n$) symmetry breaking at any finite temperature because of the proliferation of gapless excitations \cite{Mermin1966, Polyakov1975a}. Therefore, it is widely believed that there are neither extraordinary nor special transitions in this case \cite{Binder1983phase, Diehl1986phase} \footnote{However, cf. Refs. \cite{Deng2005} and \cite{Deng2006} for preliminary evidence of possible special transitions in the 3D classical Heisenberg and O(4) models, which were speculated to be Kosterlitz-Thouless transitions.}.

\begin{figure}[!bt]
\includegraphics[width=\linewidth]{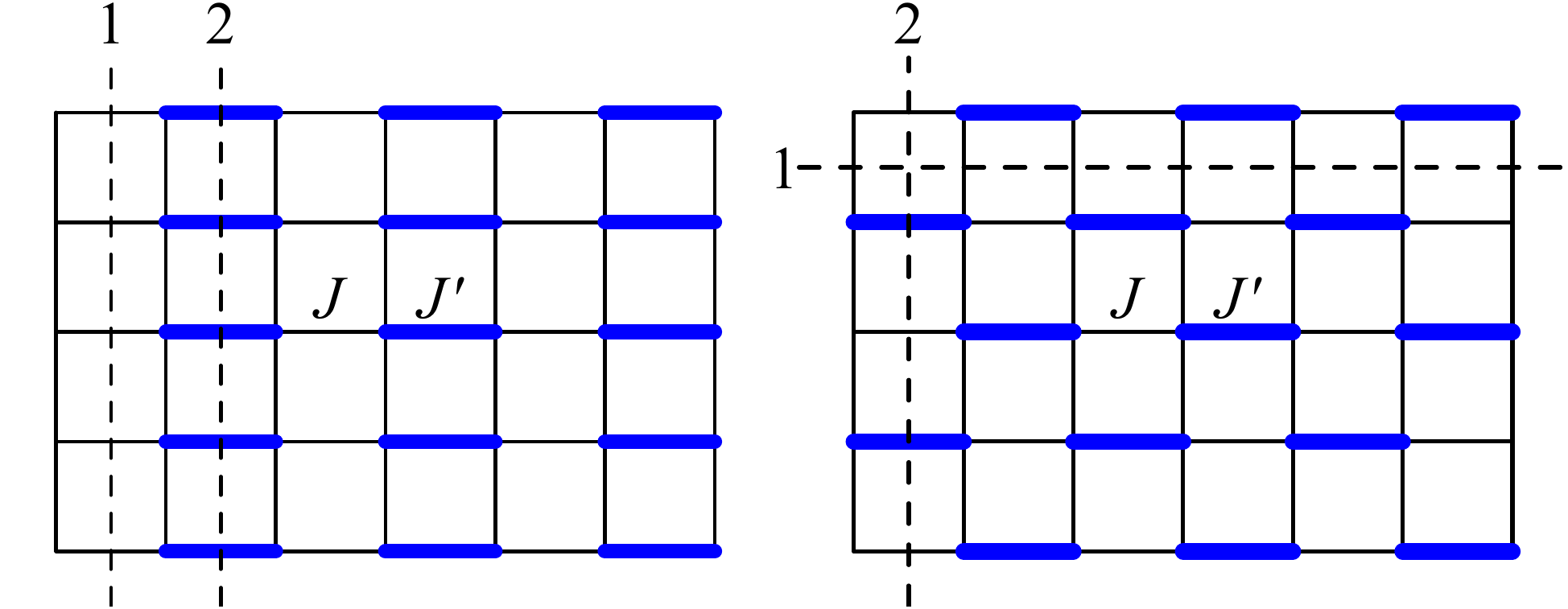}
\caption{Columnar (left) and staggered (right) dimerized spin-$1/2$ Heisenberg models on the square lattice. $J$ and $J'$ are the exchange coupling strengths on the two types of bonds, and $J'>J$. In each model, two types of open boundaries (denoted by cut-1 and 2) cutting along the two dashed lines are considered in this work.}
\label{fig:lattice}
\end{figure}

SCB also sets in at quantum critical points (QCPs) \cite{Grover2012a, Zhang2017}. In this work, we study the SCB of the dimerized spin-$1/2$ antiferromagnetic (AF) Heisenberg models on the square lattice (Fig. \ref{fig:lattice}). These models host (2+1)D O(3) QCPs between the gapped dimerized phases and the N\'eel ordered phases \cite{Matsumoto2001, Wenzel2008}. We realize all three types of SCB of 3D O(3) universality class in these models with different surface configurations.

First, we show that gapped surface states in the bulk disordered phase generally result in the ordinary transition of the 3D O(3) class at the bulk QCP. This is consistent with previous works on one of the QCPs of the decorated square lattice \cite{Zhang2017}.

Second, we show that when the bulk disordered phase has gapless surface states, the SCB belongs to a universality class different from the ordinary transition. This SCB universality class was first discovered in the decorated square lattice Heisenberg model and was taken as a feature of the symmetry-protected topological (SPT) order \cite{Zhang2017}. In the present work, the same SCB universality recurs in the columnar model with the surface cut-2 [Fig. \ref{fig:lattice} (a)], where the surface spins form an AF Heisenberg chain and are gapless in the bulk disordered phase. It indicates that this SCB universality class is a generic consequence of gapless surface states. The coexistence of the surface and the bulk critical states at the QCP suggests that this SCB universality class corresponds to the multicritical special transition of the 3D O(3) class, even though the latter is precluded in classical phase transitions.

Third, an extraordinary transition is realized in the staggered dimerized model with the surface cut-2 [Fig. \ref{fig:lattice} (b)]. A ferrimagnetic order forms on the surface both in the bulk disordered phase and at the QCP. The surface critical exponents at the extraordinary transition are found to be $\eta_{\parallel} = 1.004(13)$ and $\eta_{\perp} = -0.5050(10)$, which are inconsistent with previous theoretical predictions based on a general scaling theory and the large-$n$ expansion \cite{Bray1977, Ohno1984}, and even violate the scaling relation in Eq. (\ref{eq:etas}). Therefore, the theory of extraordinary transitions must be substantially modified to account for our finding.
 
\emph{Models and Method.---}In this work, we study the SCB of the dimerized spin-$1/2$ Heisenberg models on the square lattice (Fig. \ref{fig:lattice}). The Hamiltonians are given by
\begin{equation}
H=J\sum\limits_{\langle i,j\rangle}\mathbf{S}_i\cdot\mathbf{S}_j+J'\sum\limits_{\langle i,j\rangle'}\mathbf{S}_i\cdot\mathbf{S}_j,
\end{equation}
in which $J$ and $J'$ are the coupling strengths of the weak and the strong bonds (denoted by thin and thick lines), respectively. The strong bonds either form a columnar pattern [Fig. \ref{fig:lattice} (a)] or a staggered pattern [Fig. \ref{fig:lattice} (b)], which are called the columnar model and the staggered model, respectively.

In both models, the ground state has long-range N\'eel order if $J'/J \simeq 1$. For $J'/J \gg 1$, the ground state is adiabatically connected to the direct product state of the spin singlets on the strong bonds, thus is disordered with a nonzero energy gap. Previous studies have unveiled a continuous quantum phase transition from the disordered phase to the N\'eel ordered phase in each model \cite{Matsumoto2001, Wenzel2008}. The QCP of the columnar model lies at $J'/J = 1.9096(4)$, and unambiguously belongs to the 3D O(3) universality class \cite{Matsumoto2001}. The QCP of the staggered model at $J'/J = 2.5196(2)$ is more controversial. The first numerical simulation found the critical exponents to be $\nu = 0.689(5)$ and $\eta = 0.09(1)$, which significantly deviate from the 3D O(3) universality class \cite{Wenzel2008}. However, this conclusion was challenged in later works and the deviation was attributed to strong irrelevant corrections \cite{Fritz2011, Jiang2012c, Ma2018}.

In this work, we study the SCB of both models. We use the periodic boundary condition along one direction and the open boundary condition along the other direction to expose the surface. Two different surface configurations are considered in each model, which cut along the dashed lines shown in Fig. \ref{fig:lattice} (denoted by cut-1 and 2, respectively).

The projective quantum Monte Carlo algorithm in the valence bond basis \cite{Sandvik2005, Sandvik2010a} is adopted. The calculations are performed at the QCPs unless stated otherwise. The lattice size is $L\times L$, with $8\leq L\leq 80$. $10^{7}$ Monte Carlo sweeps are performed for each surface configuration.

The squared staggered magnetization of the surface spins $m_{s1}^{2}$ and the spin correlation functions $C_{\parallel}(L/2)$ and $C_{\perp}(L/2)$ are adopted to characterize the SCB. $C_{\parallel}(r)$ and $C_{\perp}(r)$ are equal-time spin correlation functions with one point fixed on the surface and the other point moving parallel ($C_{\parallel})$ or perpendicular to ($C_{\perp}$) the surface. They obey the following finite size scaling forms \cite{Binder1974},
\begin{align}
m_{s1}^{2}\cdot L &=c + L^{2y_{h1}-3}(b_{0}+b_{1} L^{y_i}), \label{eq:yh1fss}\\
|C_{\parallel}(L/2)| &=L^{-1-\eta_{\parallel}}(b_{0}+b_{1} L^{y_i}), \label{eq:etaparafss}\\
|C_{\perp}(L/2)| &=L^{-1-\eta_{\perp}}(b_{0}+b_{1} L^{y_i}), \label{eq:etaperpfss}
\end{align}
in which $y_{h1}$ is the scaling dimension of the surface staggered magnetic field $h_{1}$, and $\eta_{\parallel}$ and $\eta_{\perp}$ are the surface anomalous dimensions. The constant term $c$ in Eq. (\ref{eq:yh1fss}) encodes the short-range nonuniversal contribution to $m_{s1}^{2}$. $b_{i}$'s are nonuniversal fitting parameters. $y_{i}$ is the irrelevant correction exponent. In practice, we find that setting $y_{i}=-1$ yields good fitting to all numerical results.

The critical exponents $y_{h1}$, $\eta_{\parallel}$ and $\eta_{\perp}$ are expected to obey the following relations \cite{Barber1973, Lubensky1975},
\begin{align}
\eta_{\parallel} &= d-2y_{h1}, \label{eq:etayh1}\\
2\eta_{\perp} &= \eta_{\parallel} + \eta \label{eq:etas},
\end{align}
in which $d=3$ is the spacetime dimension, and $\eta$ is the bulk anomalous dimension. These relations serve as consistency check to our simulations.

\begin{figure}[!bt]
\includegraphics[width=\linewidth]{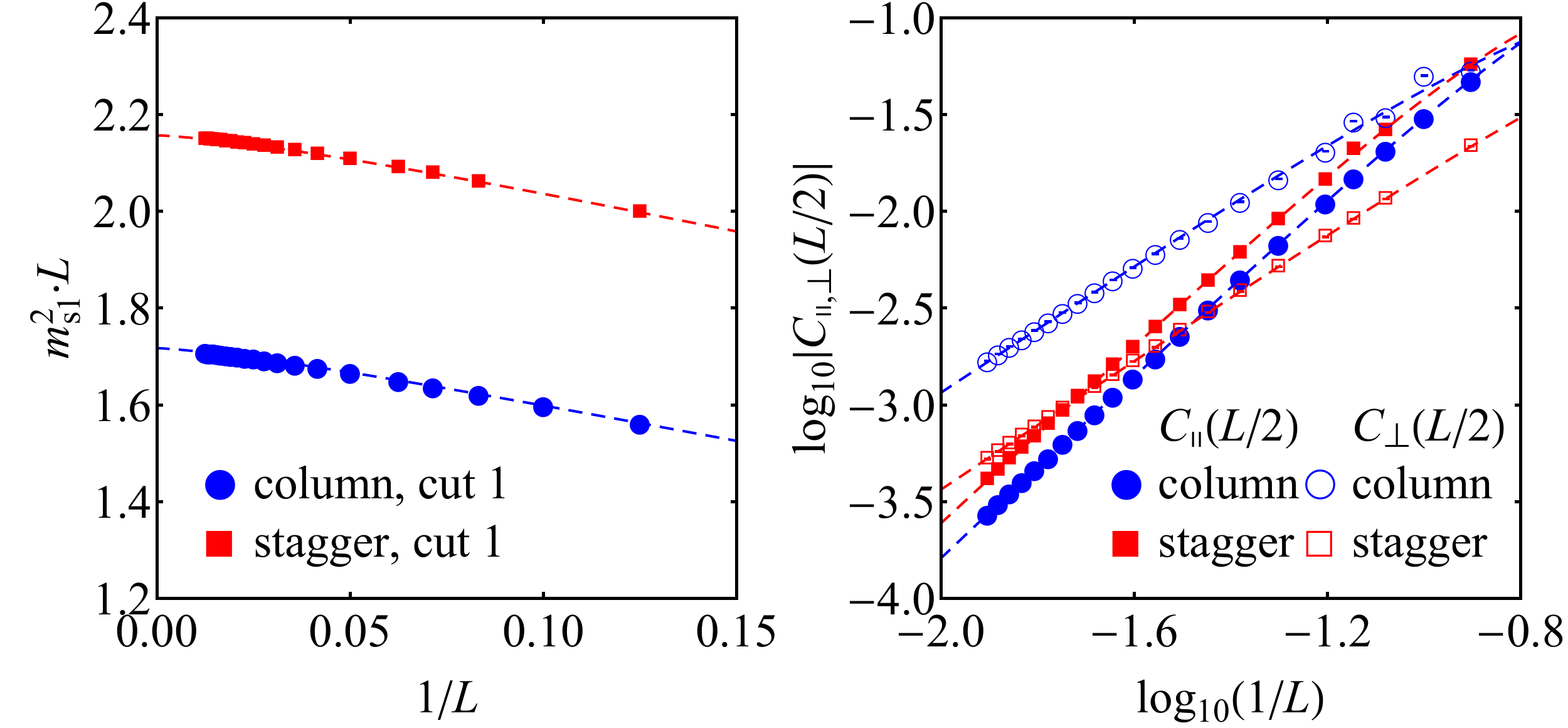}
\caption{Physical quantities at the ordinary transitions in surface cut-1 of the columnar and the staggered models: $m_{s1}^{2}\cdot L$ (left panel), and $C_{\parallel}(L/2)$ and $C_{\perp}(L/2)$ (right panel). The dashed lines are the finite-size scaling functions.}
\label{fig:ord}
\end{figure}

\begin{ruledtabular}
\begin{table}[!b]
\caption{Surface critical exponents of the dimerized Heisenberg models with different surface cut configurations. Results of the decorated square lattice at the trivial phase-N\'eel QCP ($J_{c1}$) and the AKLT-N\'eel QCP ($J_{c2}$) \cite{Zhang2017}, the 3D classical Heisenberg model \cite{Deng2005}, and the field theoretic results for the ordinary (ord.) and the special (sp.) transitions from various techniques, including $\epsilon=4-d$ expansion \cite{Diehl1980, Diehl1981}, $\epsilon=d-2$ expansion \cite{Diehl1986}, massive field theory \cite{Diehl1994, Diehl1998} and conformal bootstrap \cite{Gliozzi2015}, and the anomalous dimensions of transverse (trans.) and longitudinal (long.) correlations from the scaling arguments and the large-$n$ expansion of O($n$) models at the extraordinary (ext.) transition \cite{Bray1977, Ohno1984} are also listed for comparison.}
\label{tab:exponents}
\begin{tabular}{c c c c c}
Class			&	Model									&	$y_{h1}$	&	$\eta_\parallel$	&	$\eta_\perp$	\\
\hline
Ord.			&	Column, cut-1							&	$0.840(17)$	&	$1.387(4)$			&	$0.67(6)$		\\
				&	Stagger, cut-1							&	$0.830(11)$	&	$1.340(21)$			&	$0.682(2)$		\\
				&	Deco.sq., $J_{c1}$						&	$0.810(20)$	&	$1.327(25)$			&	$0.680(8)$		\\
				&	3D classical							&	$0.813(2)$												\\
				&	$\epsilon=4-d$ exp.						&	$0.846$		&	$1.307$				&	$0.664$			\\
				&	$\epsilon=d-2$ exp.						&				&	$1.39(2)$			&					\\
				&	Massive field							&	$0.831$	&	$1.338$				&	$0.685$			\\
				&	Bootstrap								&	$0.831$		&						&					\\
\hline
Sp.				&	Column, cut-2							&	$1.7339(12)$&	$-0.445(15)$		&	$-0.218(8)$		\\
				&	Deco.sq., $J_{c2}$						&	$1.7276(14)$&	$-0.449(5)$			&	$-0.2090(15)$	\\
				&	$\epsilon=4-d$ exp.						&	$1.723$		&	$-0.445$			&	$-0.212$		\\
\hline
Ext.			&	Stagger, cut-2							&				&	$1.004(13)$			&	$-0.5050(10)$	\\
				&	Scaling, trans.							&				&	$3$					&	$3/2$	\\
				&	Scaling, long.							&				&	$5$					&	$(5+\eta)/2$	\\
\end{tabular}
\end{table}
\end{ruledtabular}

\emph{Ordinary transition.---}The surface cut-1 in both models do not break any strong bonds, thus the surface states remain gapped in the bulk disordered phases. The power-law correlation on the surface at the QCP is purely induced by the critical bulk states.

The numerical results of $m_{s1}^{2}$, $C_{\parallel}(L/2)$ and $C_{\perp}(L/2)$ are shown in Fig. \ref{fig:ord}. In the columnar model with surface cut-1, the finite-size scaling yields $y_{h1}=0.840(17)$, $\eta_{\parallel}=1.387(4)$ and $\eta_{\perp}= 0.67(6)$. Similar analysis on the staggered model with surface cut-1 gives $y_{h1}=0.830(11)$, $\eta_{\parallel}=1.340(21)$ and $\eta_{\perp}=0.682(2)$. These critical exponents are listed in Table \ref{tab:exponents}. All of them obey the relations in Eqs. (\ref{eq:etayh1}) and (\ref{eq:etas}), and are consistent with the ordinary transition of the 3D O(3) class. This is not a surprise for the columnar model, where the 3D O(3) universality class of the bulk QCP has been well-established \cite{Matsumoto2001, Wenzel2008}. For the staggered model, this SCB universality implies that the bulk QCP also belongs to the 3D O(3) class, which may help to resolve the controversy \cite{Wenzel2008, Fritz2011, Jiang2012c}.

\begin{figure}[!bt]
\includegraphics[width=\linewidth]{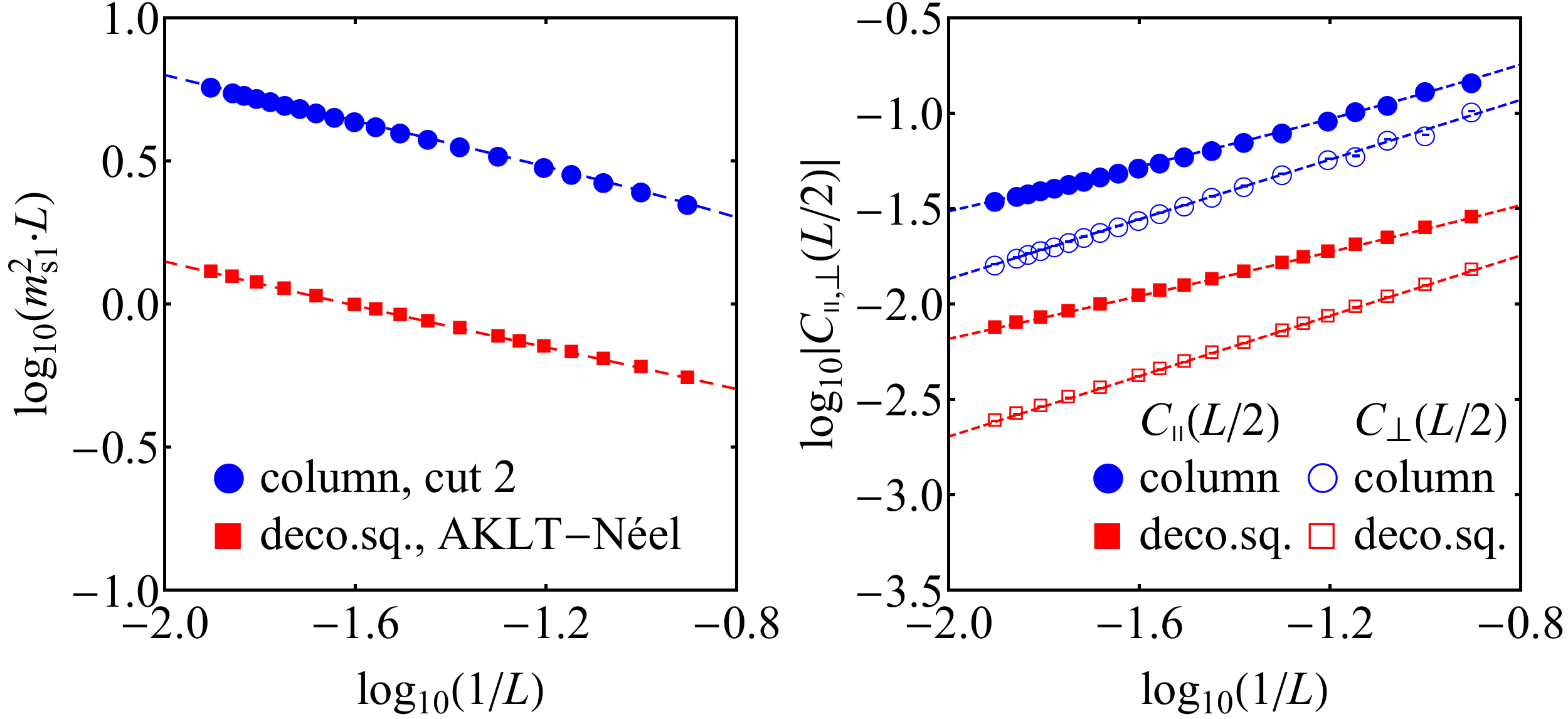}
\caption{Physical quantities at the special transition of the columnar model with surface cut-2, and the AKLT-N\'eel QCP of the decorated square lattice model \cite{Zhang2017}: $m_{s1}^{2}\cdot L$ (left panel) and $C_{\parallel,\perp}(L/2)$ (right panel).}
\label{fig:special}
\end{figure}

\emph{Special transition.---}The surface cut-2 of the columnar model breaks the strong bonds and leaves dangling bonds on the surface. In the bulk disordered phase, these dangling bonds form a spin-$1/2$ Heisenberg chain with short-range AF coupling, which is gapless according to the Lieb-Schultz-Mattis theorem \cite{Lieb1961}. This is similar to the emergence of a gapless surface state in the symmetry-protected topological (SPT) Affleck-Kennedy-Lieb-Tasaki (AKLT) phase \cite{Affleck1987a, Chen2011b, Takayoshi2016a, Zhang2017}, even though the bulk of the columnar model is not an AKLT phase. The engineering of gapless surface states with dangling spins was also studied in Ref. \cite{Suzuki2012b}.

The QCP from the AKLT to the N\'eel ordered phase was studied in Ref. \cite{Zhang2017}, and the SCB was shown to be in a distinct universality class from the ordinary transition. This was attributed to the interaction of the gapless surface state of the SPT phase and the critical bulk state \cite{Zhang2017}, and was later interpreted as a gapless SPT state \cite{Scaffidi2017, Parker2017}.

The gapless surface state of the columnar model with cut-2 results in the same SCB at the QCP as the AKLT-N\'eel transition. This is evident from the numerical results shown in Fig. \ref{fig:special}. The critical exponents of the columnar model with cut-2 from the finite-size scaling are given by $y_{h1}=1.7339(12)$, $\eta_{\parallel}=-0.445(15)$ and $\eta_{\perp}= -0.218(8)$, which are consistent with those of the AKLT-N\'eel transition. Therefore, this SCB class is a general consequence of the coexistence of the critical states both in the bulk and on the surface.

The coexistence of the surface and the bulk critical states suggests that this SCB universality class is the the multicritical special transition of the 3D O(3) model. In the field theoretic approach to the SCB of $d$-dimensional O($n$) models, the surface critical exponents of the special transition were calculated with $\epsilon$-expansion ($\epsilon=4-d$) up to the $\epsilon^{2}$ order, e.g., $\eta_{\parallel}$ is given by \cite{Diehl1981, Diehl1986phase}
\begin{equation}
\eta_{\parallel} = -\frac{n+2}{n+8}\epsilon +\frac{5(n+2)(4-n)}{2(n+8)^{2}}\epsilon^{2}.
\end{equation}
Setting $\epsilon=1$ and $n=3$ yields $\eta_{\parallel} = -0.445$, which is (quite unexpectedly) consistent with our numerical results. Other surface critical exponents are obtained similarly and are listed in Table \ref{tab:exponents} for comparison.

We remark that the special transition was never anticipated in the 3D classical O(3) model \cite{Binder1983phase, Diehl1986phase} because the 2D surface cannot possess either long-range order or power-law correlation at any finite temperature. The critical surface states of the columnar model with cut-2 and the AKLT phase are of pure quantum origin: the topological $\theta$-term in the effective field theory of the spin-$1/2$ AF Heisenberg chain suppresses the topological defects and leads to a critical state at the ground state \cite{Haldane1985, Affleck1986}. In contrast, the proliferation of these defects renders the 2D surface of the 3D classical O(3) model always short-range correlated. Moreover, the robustness of the critical surface states leaves the special transitions in both models less fine-tuned, i.e.,  unlike the 3D classical Ising model (Fig. \ref{fig:phasediag}), the special transitions naturally occur at the bulk QCPs without tuning the surface coupling strength.

\begin{figure}[!bt]
\includegraphics[width=\linewidth]{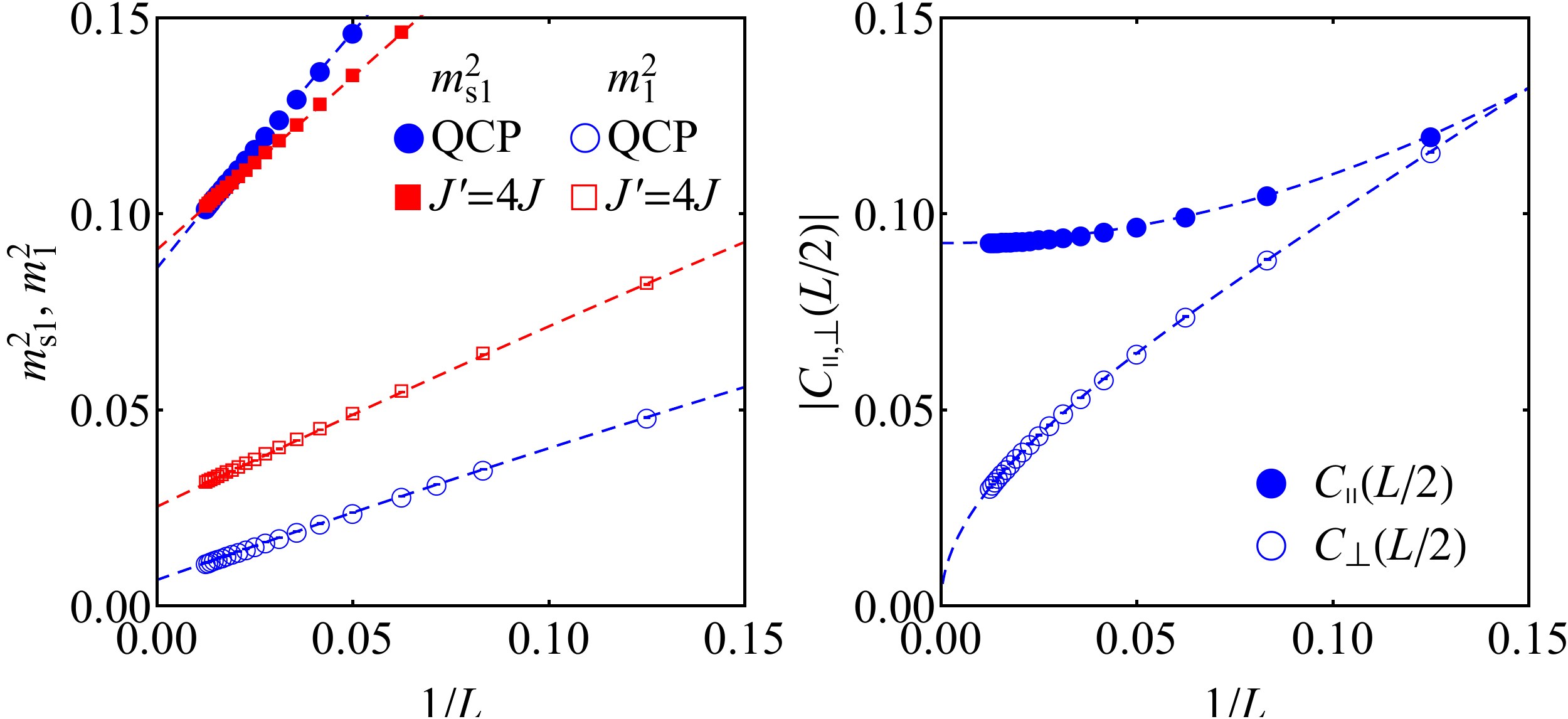}
\caption{Surface physical quantities of the staggered model with cut-2. Left: The staggered and the uniform magnetizations on the surface, $m_{s1}^{2}$ and $m_{1}^{2}$, extrapolate to nonzero values as $L\rightarrow \infty$ both in the bulk disordered phase ($J'=4J$) and at the QCP, revealing the FI order on the surface. Right: $C_{\parallel}(L/2)$ and $C_{\perp}(L/2)$ at the QCP.}
\label{fig:extra}
\end{figure}

\emph{Extraordinary transition.---}In the staggered model, the surface cut-2 exposes a surface with two inequivalent sublattices [Fig. \ref{fig:lattice} (b)]. In the bulk dimerized phase, the extensive degeneracy of the dangling bonds is lifted by their effective ferromagnetic (FM) coupling. A long-range FM order sets in on this sublattice at the ground state. The AF coupling to the other sublattice induces a weaker antiparallel magnetization on the other sublattice. Therefore, there is a ferrimagnetic (FI) order on the surface at the ground state.

The squared uniform and staggered magnetizations on the surface, $m_{1}^{2}$ and $m_{s1}^{2}$, are shown in Fig. \ref{fig:extra} (a). All these surface order parameters extrapolate to nonzero values in the thermodynamic limit both in the bulk disordered phase (taking $J'=4J$ as an example) and at the QCP. The preformed surface FI order indicates that the QCP is an extraordinary transition.

In order to study the extraordinary transition, one must single out the surface singularities induced by the bulk QCP. However, this is very difficult for thermodynamic quantities even in the mean field theory and exactly solvable models \cite{Lubensky1975a, Bray1977}, because the singularities at the extraordinary transitions are so weak that they are often overshadowed by nonsingular contributions. Therefore, we restrict our attention to the spin correlations $C_{\parallel}(L/2)$ and $C_{\perp}(L/2)$ at the QCP, which are shown in Fig. \ref{fig:extra} (b). $C_{\parallel}(L/2)$ decreases with a power law and saturates at a nonzero value as $L\rightarrow \infty$ due to the surface FI order, i.e., $C_{\parallel}(L/2)=c+aL^{-(1+\eta_{\parallel})}$, and $\eta_{\parallel}=1.004(13)$. This behavior is distinct from the FI spin chains, where the spin correlations drop exponentially \cite{Ivanov1998, Wu1999}, hence it is induced by the bulk critical state, and captures the surface singularity at the extraordinary transition. On the other hand, $C_{\perp}(L/2)$ follows a pure power-law decay, $C_{\perp}(L/2)=aL^{-(1+\eta_{\perp})}$ with $\eta_{\perp}=-0.5050(10)$.

These anomalous dimensions are inconsistent with theoretical predictions for the extraordinary transitions of $d$-dimensional O($n$) models based on scaling arguments and large-$n$ expansion \cite{Bray1977, Ohno1984} listed in Table \ref{tab:exponents}. Moreover, they apparently violate the relation in Eq. (\ref{eq:etas}). This remarkable feature suggests that the general understanding of extraordinary transitions based on scaling theory \cite{Bray1977} is incomplete. First, the surface order may induce a different length scale besides the bulk correlation length, and thus invalidate the simple scaling arguments. Second, the violation of the scaling relation in Eq. (\ref{eq:etas}) may be attributed to the dichotomy of the transverse and the longitudinal correlations in the presence of the surface order, i.e., while $C_{\parallel}(r)$ is usually dominated by the transverse correlation, $C_{\perp}(r)$ may be mainly contributed by the longitudinal correlation. These possibilities must be examined by further calculations.

\emph{Summary.---}The surface critical behavior (SCB) of two dimerized Heisenberg models at their bulk quantum critical points are studied with different surface cut configurations. We show that all three types of SCB, i.e., the ordinary, special and extraordinary transitions of the 3D O(3) model, are realized with certain surface configurations. Gapped surface states in the bulk disordered phase generally lead to ordinary transitions, and gapless surface states generally result into multicritical special transitions even if the latter is precluded in the 3D classical O(3) models. We also find a ferrimagnetic order on the surface cut-2 of the staggered model, which leads to an extraordinary transition. The surface anomalous dimensions $\eta_{\parallel}$ and $\eta_{\perp}$ at this extraordinary transition not only contradict previous theoretical predictions based on scaling arguments and violate the scaling relation. This feature poses a serious challenge to our current understanding of extraordinary transitions.

\emph{Note added.---}After the submission of this work, an independent work \cite{Weber2018} was posted in arXiv, in which the authors also numerically studied the generic correspondence between the ordinary and special classes of surface critical behavior and the different types of surface states. Their results are fully consistent with our work.

\acknowledgments
We thank Youjin Deng for interesting discussions, and Masahiro Sato and Hans W. Diehl for bringing Ref. \cite{Suzuki2012b} and Refs. \cite{Diehl1986, Diehl1994, Diehl1998, Gliozzi2015} to our attention. This work is supported by the National Natural Science Foundation of China under Grant No. 11774002 (C. Ding), 11775021 and 11734002 (W. Guo) and the Anhui Provincial Natural Science Foundation under Grant No. 1508085QA05 (C. Ding). L. Zhang is supported by the Key Research Program of the Chinese Academy of Sciences (Grant No. XDPB08-4) and the start-up funding of UCAS.

\bibliography{/Dropbox/ResearchNotes/Bibtex/library,/Dropbox/ResearchNotes/Bibtex/Books}

\end{document}